\documentclass[journal=jacsat,manuscript=article]{achemso}

\usepackage[version=3]{mhchem} 

\usepackage[version=3]{mhchem} 
\usepackage{physics}
\usepackage{parskip} 
\usepackage[utf8]{inputenc}
\usepackage{subcaption}
\usepackage[toc,page]{appendix}
 \SectionNumbersOn

\author{Rishabh Gupta}
\affiliation[Purdue University]
{Department of Chemistry, Purdue University, West Lafayette, IN, USA}

\author{Sabre Kais}
\email{kais@purdue.edu}
\affiliation[Purdue University]
{Department of Chemistry, Department of Physics and Astronomy, and Purdue Quantum Science and Engineering Institute, Purdue University, West Lafayette, IN, USA}

\author{Raphael D. Levine}
\email{raphy@mail.huji.ac.il}
\affiliation[Hebrew University]
{The Fritz Haber Center for Molecular Dynamics and Institute of Chemistry, The Hebrew University of Jerusalem, Jerusalem 91904, Israel
}
\alsoaffiliation[University of California]
{Department of Chemistry and Biochemistry and  Department of Molecular and
Medical Pharmacology, David Geffen School of Medicine, University of California, Los
Angeles, CA 90095, USA}

\title[An \textsf{achemso} demo]
  {Convergence of reconstructed density matrix to a pure state using maximal entropy approach}

\setcitestyle{numbers}
\mciteErrorOnUnknownfalse
\DeclareUnicodeCharacter{2212}{-}
\begin{document}







\begin{abstract}
Impressive progress has been made in the past decade in the study of technological applications of varied types of quantum systems. With industry giants like IBM laying down their roadmap for scalable quantum devices with more than 1000-qubits by the end of 2023, efficient validation techniques are also being developed for testing quantum processing on these devices. The characterization of a quantum state is done by experimental measurements through the process called quantum state tomography (QST) which scales exponentially with the size of the system. However, QST performed using incomplete measurements is aptly suited for characterizing these quantum technologies especially with the current nature of noisy intermediate-scale quantum (NISQ) devices where not all mean measurements are available with high fidelity.  We, hereby, propose an alternative approach to QST for the complete reconstruction of the density matrix of a quantum system in a pure state for any number of qubits by applying the maximal entropy formalism on the pairwise combinations of the known mean measurements. This approach provides the best estimate of the target state when we know the complete set of observables which is the case of convergence of the reconstructed density matrix to a pure state. Our goal is to provide a practical inference of a quantum system in a pure state that can find its applications in the field of quantum error mitigation on a real quantum computer that we intend to investigate further.  
\end{abstract}

\section{Introduction}
The rapid advancements towards the development of large scale quantum computing devices in recent years require efficient methods that can validate information processing on these devices. Quantum state tomography (QST) is one such standard data-driven technique that can characterize the quantum mechanical state of the system based on the information of the expectation values of a complete set of observables [\cite{nielson},\cite{kais},\cite{photonic},\cite{qubits},\cite{cramer}]. However, the increase in the size of quantum systems poses a critical limitation on QST due to the exponential scaling of the number of parameters required to reconstruct a quantum state that is a tensor product of qubits. In practice, full QST for large quantum systems has been performed on not more than 10-qubits [\cite{10-qubit}]. \\
We are currently in the era of noisy intermediate-scale quantum (NISQ) computing [\cite{preskill2018quantum}] which restricts the use of a large quantum device for practical purposes such as recovering the true quantum state, owing to the inherent noise in the result. Measurements on these NISQ devices are therefore of limited fidelity and in certain cases we have access to only a limited number of observations. This along with the scaling problem makes QST intractable in real experiments except for small quantum systems. Various techniques have been suggested to address the underlying scaling problem as well as to mitigate against imperfect measurements [\cite{detector}]. Some of the proposed tomography methods are matrix product state tomography [\cite{cramer}], neural network tomography [\cite{Torlai2018},\cite{Carrasquilla2019},\cite{Palmieri2020},\cite{vasconcel}], quantum overlapping tomography [\cite{overlap}], shadow tomography [\cite{aaronson},\cite{Huang2020}]. Apart from the conventional tomography techniques, there have been approaches that try to estimate the quantum state based on incomplete measurements [\cite{wich},\cite{jaynes},\cite{katz1967principles}]. Some of these methods include quantum state estimation using maximum likelihood estimation [\cite{hradil},\cite{baumgratz2013scalable}], and Bayesian state estimation [\cite{lukens2020practical},\cite{lukens2020bayesian}]. \\
In our previous work [\cite{rishabh}] we proposed an alternative approach to QST based on the maximal information entropy formalism [\cite{jaynes},\cite{raphy},\cite{raphy2}] using finite but incomplete set of measurements that serve as the constraints of the problem. These are special kinds of constraints that correspond to the mean values of populations and coherences.We showed that using the maximal entropy approach we can obtain an accurate prediction of an unknown mean measurement (a probability) using a pair of known mean measurements (a probability and a coherence).  In addition to the validation of the results of quantum calculations performed on NISQ devices using QST, there are a variety of further circumstances where the mean values of populations and coherences are of primary interest [\cite{fresch},\cite{Gattuso2020},\cite{fresch2}]. With an incomplete set of known mean measurements, maximizing the von Neumann entropy [\cite{wich},\cite{Buzek_1997},\cite{alhassid},\cite{dagan}] of the system provides an additional constraint to obtain a unique solution for the state determination. In this paper, we extend this approach to reconstruct the complete density matrix of an \textit{n}-qubit quantum system with a special reference to a pure state using the expectation values of \textit{N} observables where \textit{N}=2$^n$. The unique feature of our approach is that we consider pairwise combination of the known probability with the known coherence to predict an unknown probability using the maximal entropy formalism and repeat this process until all the probabilities for all the \textit{n} qubits have been determined. Once all the probabilities have been calculated, we employ the same approach on the pairwise combinations of the probabilities to determine all the unknown coherences. The detailed description of the method is provided in Section \ref{meth}. To validate and support our theoretical proposition we conducted numerical simulations in IBM's qiskit [\cite{Qiskit}]. We also implemented our approach on IBM quantum computing chip that can be easily accessed through IBM quantum experience [\cite{SANTOS2017}]. The results in Section \ref{res} show the accuracy of our approach when we use the expectation values of observables obtained from noise-free \textit{statevector$\_$simulator} backend in Qiskit for the reconstruction of the complete density matrix for quantum systems ranging from 2-10 qubits using the proposed approach. We intend to further analyse this approach for applications in the field of quantum error correction and compare it with the available error correction codes [\cite{coles},\cite{baek2019density}].

\section{Method \label{meth}}
A unique characterization of a quantum state requires measuring the expectation values of a complete set of observables. For the state vector of a quantum state with \textit{N} entries, this complete set is defined by \textit{2N-1} independent parameters whose knowledge is required to uniquely characterize the quantum state. For example, measurements of expectation values of 16 operators are required to completely describe a 2-qubit quantum system in pure state:
\begin{eqnarray}
    &\hspace*{0.001cm}&\{\ket{1}\bra{1},\ket{2}\bra{2},\ket{3}\bra{3},\ket{4}\bra{4},(\ket{1}\bra{2},\ket{2}\bra{1}),(\ket{1}\bra{3},\ket{3}\bra{1}),(\ket{1}\bra{4},\ket{4}\bra{1}), \nonumber \\
    &\hspace*{0.001cm}&(\ket{2}\bra{3},\ket{3}\bra{2}), (\ket{2}\bra{4},\ket{4}\bra{2}), (\ket{3}\bra{4},\ket{4}\bra{3}) \} \label{basis1}    
\end{eqnarray}
The expectation values of the above operators correspond to the probabilities and coherences in the 2-qubit system [\cite{alhassid}]. The maximal entropy formalism [\cite{raphy}] seeks to determine a probability distribution that is consistent with the known average values of certain operators $\hat{f}_k$ as well as ensuring that the von Neumann entropy of the distribution be maximal. Combining it with the method of Lagrange multipliers $\lambda_k$ [\cite{jaynes},\cite{raphy2}] yields the following form of the density operator [\cite{wich},\cite{dagan}]:
\begin{eqnarray}
\hat{\rho} = \frac{1}{Z(\lambda_{1},\ldots,\lambda_{k})}\exp\{-\sum_{k}\lambda_{k}\hat{f}_{k}\} \label{rho2}
\end{eqnarray}
where $Z(\lambda_{1},\ldots,\lambda_{k})=Tr(\exp\{-\sum_{k}\lambda_{k}\hat{f}_{k}\})$ insures the normalization as $Tr(\rho)=1$. Thus, even when a complete set of observables is not available, the maximal entropy formalism provides a unique characterization of the quantum state consistent with the given constraints. In this current work, we start by reconstructing the density matrix of maximal entropy that corresponds to the case when only two measurements are available, specifically, a probability and a coherence. In general the coherence will be a complex number so it is equivalent to two Hermitian observables. Based on the maximal entropy formalism we can write the Hermitian density operator in terms of the operators corresponding to the available observables as:
\begin{eqnarray}
\hat{\rho} = \frac{1}{Z(\lambda_{11},\lambda_{12},\lambda_{21})}\exp\{-\lambda_{11}\ket{1}\bra{1}-\lambda_{12}\ket{1}\bra{2} -\lambda_{21}\ket{2}\bra{1}\} \label{lag}
\end{eqnarray}
where the Lagrange multipliers $\lambda_k$ satisfy $\lambda_{21}$ = $\lambda_{12}^{*}$ so that the density matrix is Hermitian. One can satisfy this by writing the Lagrange multiplier of the coherences in terms of an amplitude and a phase, $\lambda_{12}$ = $\abs{\lambda_{12}}\exp{i\theta_{12}}$. The details of the prediction of an unknown probability from a known probability and a coherence are presented in our previous work [\cite{rishabh}]. The operators in the exponent of Eq. (\ref{lag}) do not commute so to obtain an explicit form for the density matrix we first diagonalize the Hermitian matrix \textbf{A} corresponding to the exponent term in Eq. (\ref{lag}) and then express the density operator in terms of the eigenvalues and eigenvectors of \textbf{A}:
\begin{eqnarray}
\hat{\rho} = \frac{1}{Z(\lambda_{11},\lambda_{12},\lambda_{21})}\sum_i\exp{\epsilon_{i}}|\phi_{i}><\phi_{i}| \label{rho_op}
\end{eqnarray}
where $\{\epsilon_i,\phi_i\}$ are the eigenvalues and eigenvectors of \textbf{A} expressed as a function of the unknown Lagrange multipliers in Eq. (\ref{lag}). For a 2-qubit system the density operator, upon reconstruction from a known probability and a coherence, so that there is only one phase, was shown to take the explicit form:
\begin{eqnarray}
\hat{\rho}&=&\frac{1}{Z}\sum_{i}\exp{\epsilon_{i}}|\phi_{i}><\phi_{i}| \nonumber \\
&=&\frac{1}{Z}(|4><4|+|3><3|+(a+b)|1><1|+(\frac{a}{k_{3}^*}+\frac{b}{k_{4}^*})|1><2|  \nonumber\\
&+&(\frac{a}{k_{3}}+\frac{b}{k_{4}})|2><1|+(\frac{a}{\abs{k_{3}}^{2}}+\frac{b}{\abs{k_{4}}^{2}})|2><2|) \label{density}
\end{eqnarray}
where Z=$\sum_{i}\exp{\epsilon_{i}}$, k$_3$ = -$\frac{\epsilon_3}{\lambda_{12}^{*}}$, k$_4$ = -$\frac{\epsilon_4}{\lambda_{12}^{*}}$, a=$\frac{\abs{k_{3}}^{2}}{\sqrt{(k^2_3+1)({k_3^*}^2+1)}}\exp{\epsilon_{3}}$, and \\ b=$\frac{\abs{k_{4}}^{2}}{\sqrt{(k^2_4+1)({k_4^*}^2+1)}}\exp{\epsilon_{4}}$. Since the basis operators are orthogonal, the mean measurements correspond to the coefficients of the operators in the reconstructed density matrix. Therefore,
\begin{eqnarray}
x_{11} &=& f(\lambda_{11},\lambda_{12},\lambda_{21}) = \langle\ket{1}\bra{1}\rangle = \frac{a+b}{Z} \label{x11} \\
x_{22} &=& g(\lambda_{11},\lambda_{12},\lambda_{21}) = \langle\ket{2}\bra{2}\rangle = (\frac{a}{\abs{k_{3}}^{2}}+\frac{b}{\abs{k_{4}}^{2}})/Z \label{x22} \\
x_{12} &=& h(\lambda_{11},\lambda_{12},\lambda_{21}) =  \langle\ket{1}\bra{2}\rangle = (\frac{a}{k_{3}^*}+\frac{b}{k_{4}^*})/Z \label{x12}
\end{eqnarray}
The state defined in Eq. (\ref{density}) is a mixed state. This is to be expected since we only provided information sufficient to define a pure state of the first qubit. In our previous work, we have already established that  from Eq. (\ref{density})  the information about x$_{22}$ can be obtained using x$_{11}$ and x$_{12}$ following the determination of the unknown Lagrange multipliers in Eq. (\ref{lag}). However, in the case that the density matrix is real valued, we can also propose that the mean measurement value x$_{12}$ can be determined using the same expression if we know x$_{11}$ and x$_{22}$ as the mean measurements in Eq. (\ref{x11}-\ref{x12}) are functions of the Lagrange multipliers ($\lambda_{11}$, $\lambda_{12}$, $\lambda_{21}$) that can be determined using the expectation values of x$_{11}$ and x$_{22}$. \\
After determining x$_{12}$ using the proposed approach we can similarly determine x$_{ij}$, for real valued coherences, using the corresponding two probabilities, x$_{ii}$ and x$_{jj}$, at a time. The full density matrix of a quantum system can be deduced if we have determined all the coefficients c$_{ij}$ in the definition of density operator in Eq. (\ref{club}):
\begin{eqnarray}
\hat{\rho} = \sum_{ij} c_{ij} \ket{i}\bra{j} \label{club}
\end{eqnarray}
The coefficient c$_{ij}$ is the expectation value of the corresponding operator $\ket{i}\bra{j}$ which can be determined, if the density matrix is real valued, by applying the maximal entropy approach on the pair of known mean measurements for the operators: $\ket{i}\bra{i}$ and $\ket{j}\bra{j}$:
\begin{eqnarray}
c_{ij} = \langle\ket{i}\bra{j}\rangle \label{cij}
\end{eqnarray}
Therefore, if the density matrix is real valued, by considering all the pairwise combinations of the \textit{N} probabilities $\ket{i}\bra{i}$ and $\ket{j}\bra{j}$ and applying the maximal entropy approach over each such combination we can determine all the real valued coherences and determine the state. The number of times the maximal entropy approach is applied on the pairwise combinations of the \textit{N} probabilities to determine all the coherences is \textit{N(N-1)/2} that is the number of unknown coherences. In Section \ref{complex} we show a more general result namely that also when the coherences are complex valued, it follows from the representation $\lambda_{12}$ = $\abs{\lambda_{12}}\exp{i\theta_{12}}$ with $\theta_{21}$ = $-\theta_{12}$ that it is sufficient to know the phases of half of the coherences in order to construct the phases of the other half. We show the application of this approach first by reconstructing the density matrix for the maximally entangled 2-qubit Bell state followed by the reconstruction of a 3-qubit multipartite entangled wave function. Thereby, we propose a scheme to also reconstruct the phases of the coherences and combine it with the maximal entropy approach to reconstruct the full density matrix for a quantum system in pure state. To summarize the method, we start with a certain number of known mean measurements (for example, \textit{N} probabilities if the target state is real; 1 probability and \textit{N-1} coherences if the target state is complex) and apply the formalism of maximal entropy on each pairwise combination of the known mean measurements. Each time we consider a pair of known mean measurements and apply the maximal entropy formalism we get a mixed state with an accurate description of an unknown mean measurement that is the coefficient of the corresponding operator term in Eq. (\ref{density}). We repeat this process for all the different pairs of the known mean measurements and determine all the unknown coefficients c$_{ij}$ in Eq. (\ref{cij}) and thereby combine everything together to reconstruct the complete density matrix of the quantum system in a pure state. To verify that such a process gives us a pure state in the absence of noise as well as the known mean measurements belonging to a pure state, we calculate the entropy and trace of the square of the density matrix at each step of the process as a measure of its convergence to a pure state in the Results and discussion section. 

\subsection{Bell state}
To reconstruct the density matrix of the 2-qubit Bell state: $\ket{B}$ = $\frac{1}{\sqrt{2}}(\ket{00}+\ket{11})$, we just need to determine the amplitude of the coherences as the target state is real. Following the approach of considering two probabilities at a time and applying the maximal entropy formalism to predict a coherence and repeating this process for all the six pairwise probability combinations for Bell state, we determine each of the coherence and combine it as per Eq. (\ref{club}) to obtain the following density matrix:
\[
\rho_{rec} =  
\begin{bmatrix}
 0.5 & 0 & 0 & 0.5  \\
  0 & 0 & 0 & 0 \\
  0 & 0 & 0 & 0 \\
  0.5 & 0 & 0 & 0.5 
 \end{bmatrix} 
\]
The predicted coherences match exactly with the coherences of the original Bell state density matrix and therefore, supports our approach of reconstruction of the density matrix.
 
\subsection{Reconstructing density matrix for multipartite entangled states}
\begin{figure}[ht!]
  \centering 
  \subcaptionbox{Prediction of x$_{12}$ using maximal entropy formalism}
{\includegraphics[width=3.2in]{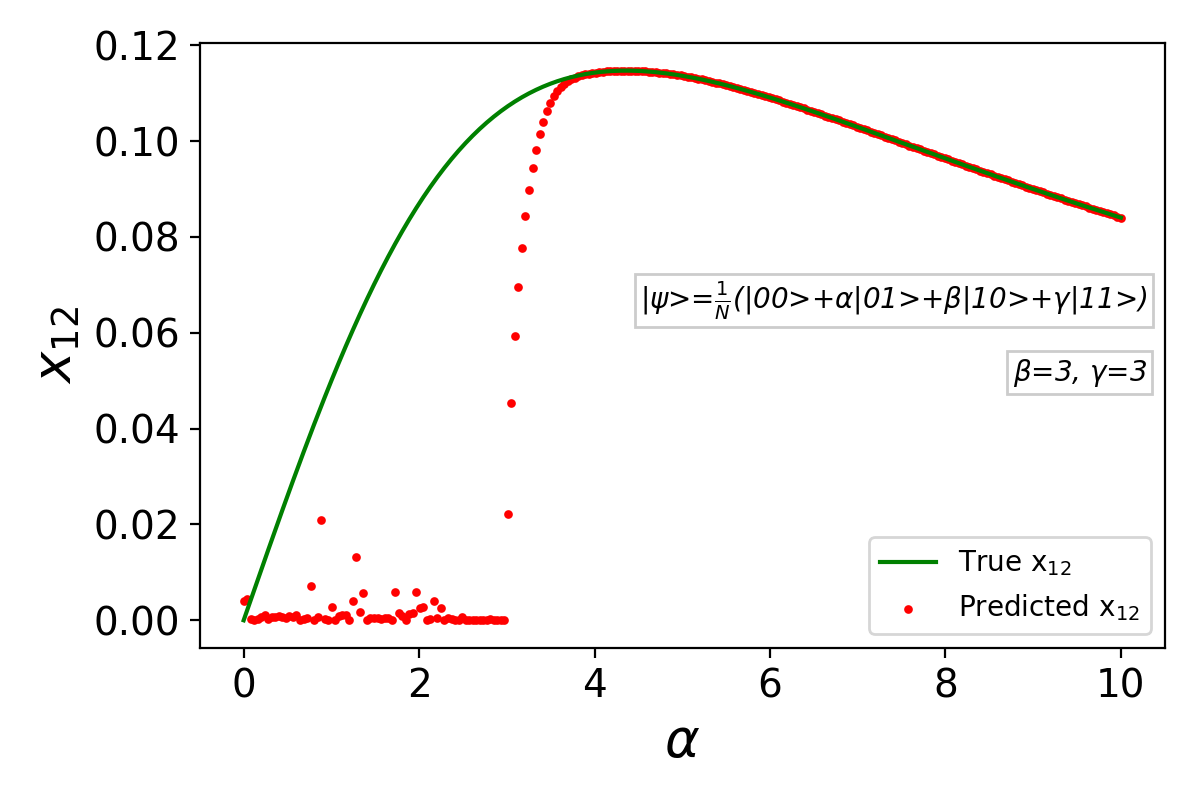}} \hspace*{\fill}
\subcaptionbox{Prediction of x$_{12}$ using maximal entropy formalism combined with scaling technique}
{\includegraphics[width=3.2in]{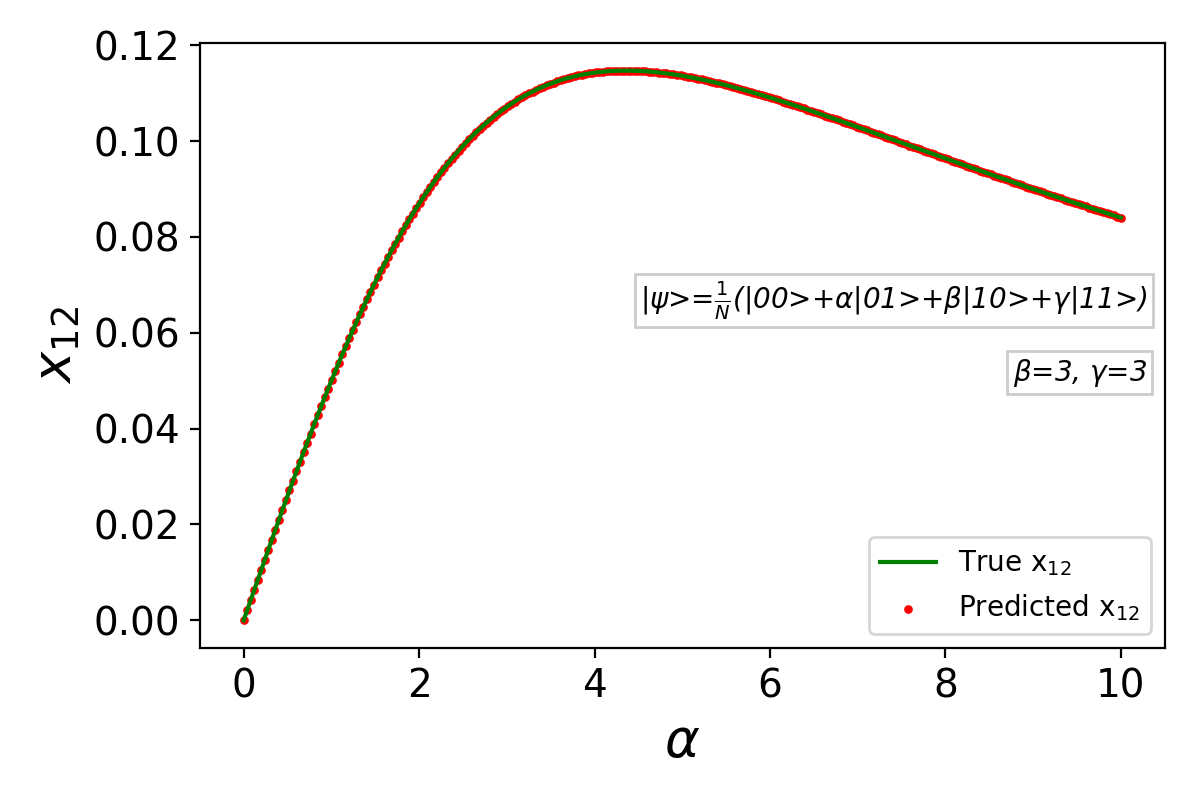}}
\caption{Plot of the predicted and true mean measurement x$_{12}$ versus the coefficient $\alpha$ of the state $\ket{01}$ using the maximal entropy formalism from: (a) two known probabilities x$_{11}$ and x$_{22}$ (b) two known probabilities x$_{11}$ and x$_{22}$ and also employing the scaling approach.  }
\label{fig_a1}
\end{figure}

Moving forwards, we consider a system with more than two non-zero coefficients in the maximally entangled state. For example, let us consider a 3-qubit entangled state: $\ket{\psi}$ = $\frac{1}{\sqrt{3}}(\ket{100} + \ket{010} + \ket{001})$. Clearly, in the true density matrix defined for this state, all the coherence terms will be zero except for the cross terms ($\langle\ket{010}\bra{100}\rangle$, $\langle\ket{001}\bra{100}\rangle$, $\langle\ket{001}\bra{010}\rangle$, and their complex conjugate) that will be 0.3333. However, when we reconstruct the density matrix as per the above approach, we obtained the following density matrix:
\[
\rho_{rec} =  
\begin{bmatrix}
  0 & 0 & 0 & 0 & 0 & 0 & 0 & 0  \\
  0 & 0.3333 & 0.3286 & 0 & 0.3286 & 0 & 0 & 0 \\
  0 & 0.3286 & 0.3333 & 0 & 0.3286 & 0 & 0 & 0 \\
  0 & 0 & 0 & 0 & 0 & 0 & 0 & 0 \\
  0 & 0.3286 & 0.3286 & 0 & 0.3333 & 0 & 0 & 0 \\
  0 & 0 & 0 & 0 & 0 & 0 & 0 & 0 \\
  0 & 0 & 0 & 0 & 0 & 0 & 0 & 0 \\
  0 & 0 & 0 & 0 & 0 & 0 & 0 & 0 
 \end{bmatrix} 
\]
We see that some errors show up in the determination of coherences using the maximal entropy approach. The predicted coherence values further diverge from the true value if we have more non-zero coefficients of the basis states defining the maximally entangled state. To further demonstrate this we consider a general 2-qubit quantum state with real amplitudes: $\ket{\psi} = \frac{1}{\sqrt{N}}(\ket{00}+\alpha\ket{01}+\beta\ket{10}+\gamma\ket{11})$ and try to predict the coherence x$_{12}$ using the probabilities x$_{11}$ and x$_{22}$ for various values of $\alpha$ and for fixed values for $\beta$ and $\gamma$. As can be seen in Figure \ref{fig_a1}a the prediction of the unknown mean measurement is highly inaccurate for smaller coefficient values of $\alpha$ and it improves considerably for larger values. This is primarily because upon solving the transcendental Eq. (\ref{x11}) and (\ref{x22}), the determined value of the Lagrange multiplier $\lambda_{12}$ is close to zero that leads to a numerical singularity while reconstructing the density matrix as $\lambda_{12}$ is in the denominator when we calculate the projection operators in the eigen basis. To address this problem we used a simple scaling approach described in detail in the Supplementary Information. We then apply the maximal entropy formalism to determine the unknown coherence. The maximal entropy approach is invariant under this scaling technique and it is done to make sure that the trace of the scaled pairwise density matrix is unity. Figure \ref{fig_a1}b shows the accuracy of the prediction of the coherence using the maximal entropy formalism combined with the scaling technique for a general 2-qubit quantum system.

This approach can be followed to reconstruct the full density matrix of a quantum system with any number of qubits. In case of quantum states with real coherences, for the prediction of an unknown coherence from two known probabilities, we just need to determine the unknown Lagrange multipliers ($\lambda_{11}$, $\lambda_{12}$, $\lambda_{21}$) and apply this approach. We repeat this procedure to consider all possible pairwise combinations of probabilities and predict all the unknown coherences. Once we determine all the unknowns, we combine everything together to obtain the reconstructed density matrix given by Eq. (\ref{club}). Therefore, we can reconstruct the complete density matrix of an \textit{N}-qubit quantum system in case of real coherences from \textit{N} probabilities.

\subsubsection{Complex coherence and phase estimation \label{complex}}
So far we have established that the maximal entropy formalism combined with the scaling technique accurately predicts the amplitude of coherence from the known probabilities. However, in the case of complex coherence we need more information than just the probabilities as we also need to determine the phase of coherence which is specific to a quantum system. To estimate the phase of every coherence term in the reconstruction of the density matrix, the following phase estimation algorithm is employed. To demonstrate it, let us consider a 2-qubit quantum system defined by: 
\begin{eqnarray}
\ket{\psi} = \frac{1}{\sqrt{N}}(\alpha\exp{i\theta_1}\ket{00}+\beta\exp{i\theta_2}\ket{01}+\gamma\exp{i\theta_3}\ket{10}+\delta\exp{i\theta_4}\ket{11})
\end{eqnarray}
where $\alpha,\beta,\gamma,\delta$ are real coefficients. The coherences are then defined as:
\begin{eqnarray}
x_{12} = \langle\ket{00}\bra{01}\rangle =  \alpha\beta\exp{i(\theta_1-\theta_2)} = \alpha\beta\exp{ip_{12}} \nonumber \\
x_{13} = \langle\ket{00}\bra{10}\rangle =  \alpha\gamma\exp{i(\theta_1-\theta_3)} = \alpha\gamma\exp{ip_{13}} \nonumber \\
x_{14} = \langle\ket{00}\bra{11}\rangle =  \alpha\delta\exp{i(\theta_1-\theta_4)} = \alpha\delta\exp{ip_{14}}
\end{eqnarray}
and so on. The $p_{12}, p_{13}, p_{14}$ correspond to the phases of the coherences x$_{12}$, x$_{13}$, x$_{14}$:
\begin{eqnarray}
p_{12} = \theta_1 - \theta_2 \nonumber \\
p_{13} = \theta_1 - \theta_3 \nonumber \\
p_{14} = \theta_1 - \theta_4 
\end{eqnarray}
Now, the phase of the remaining coherence terms of the density matrix can be estimated if we have information about the phases $p_{12}, p_{13}, p_{14}$:
\begin{eqnarray}
p_{23} = \theta_2 - \theta_3 = p_{13} - p_{12} \nonumber \\
p_{24} = \theta_2 - \theta_4 = p_{14} - p_{12} \nonumber \\
p_{34} = \theta_3 - \theta_4 = p_{14} - p_{13} \label{phase}
\end{eqnarray}
Thus, we can determine the phases of all the remaining coherences using the above approach and then combine it with the determined amplitudes from maximal entropy approach to reconstruct the entire density matrix. \\
Therefore, in case of complex coherences, for the prediction of an unknown coherence from two known probabilities, apart from the determination of the unknown Lagrange multipliers ($\lambda_{11}$, $\lambda_{12}$, $\lambda_{21}$), we also need to determine the phase ($\theta$) of coherence. For determining the phase of all the coherence terms of the density matrix, we need information about the phase of \textit{N-1} coherence terms as seen in Eq. (\ref{phase}). In our previous work we have shown that we can accurately predict an unknown probability from a known probability and a coherence. Therefore, given \textit{N-1} coherences and a probability we can construct all the probabilities using the pairwise combination of the known probability with the known coherences and applying the maximal entropy approach. After the determination of all the probabilities, we take the pairwise combination of all the probabilities and using the above approach for phase estimation and amplitude determination using maximal entropy formalism, we construct all the unknown coherences and combine everything to obtain the complete density matrix for a general pure state with complex amplitudes from \textit{N} observables (\textit{N-1} coherences and \textit{1} probability).

\section{Results and discussion \label{res}}
In this current work we propose to reconstruct the complete density matrix for a general pure state with complex amplitudes from \textit{N} observables (\textit{N-1} coherences and \textit{1} probability) using the method discussed in Section \ref{meth}. The approach that we follow can work for a quantum system with any number of qubits. To test and validate the proposed theory, we conducted numerical experiments in IBM's Qiskit [\cite{Qiskit}] which is an open-source quantum computing platform with prototype quantum devices to run and simulate quantum programs. We also implemented the theory on IBM's quantum computing chip [\cite{ibm}] and used the measurement data to reconstruct the density matrix and then compare it with the true result.  

\subsection{Trace distance and fidelity}
\begin{figure}[ht!]
  \centering 
\includegraphics[width=3.2in]{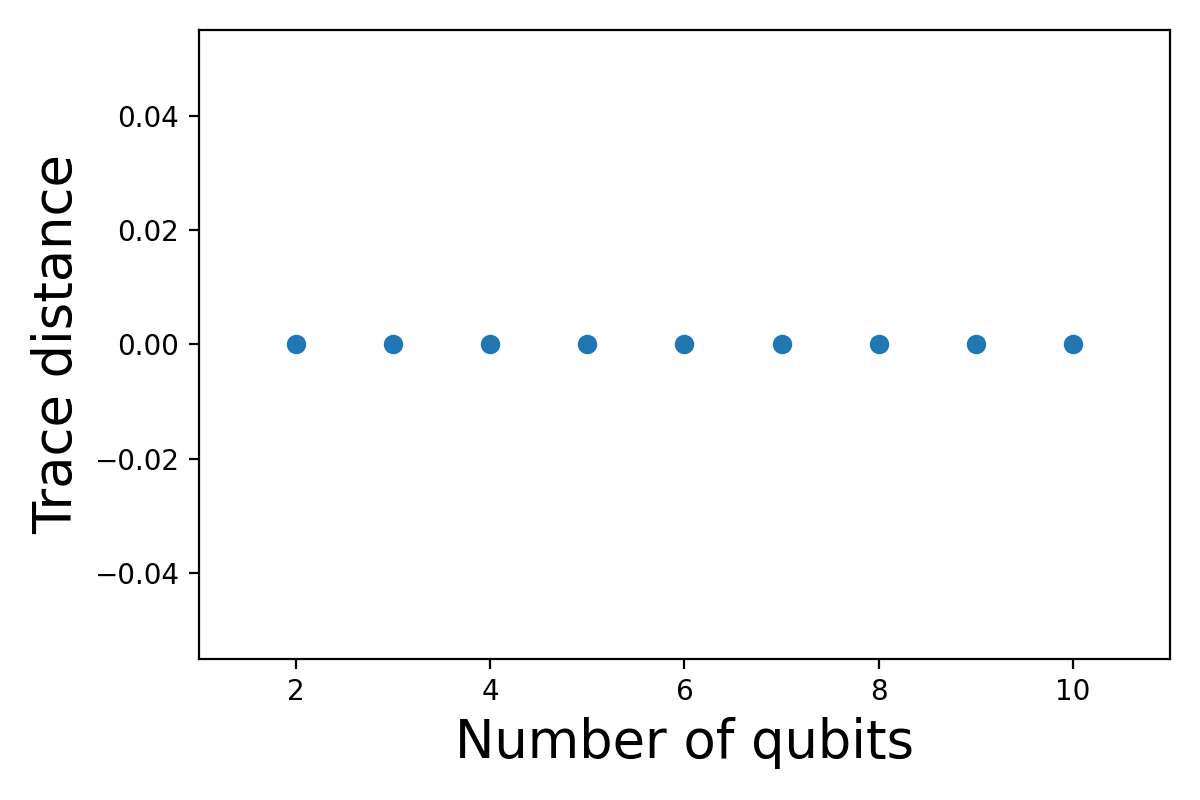} \hspace*{\fill}
\includegraphics[width=3.2in]{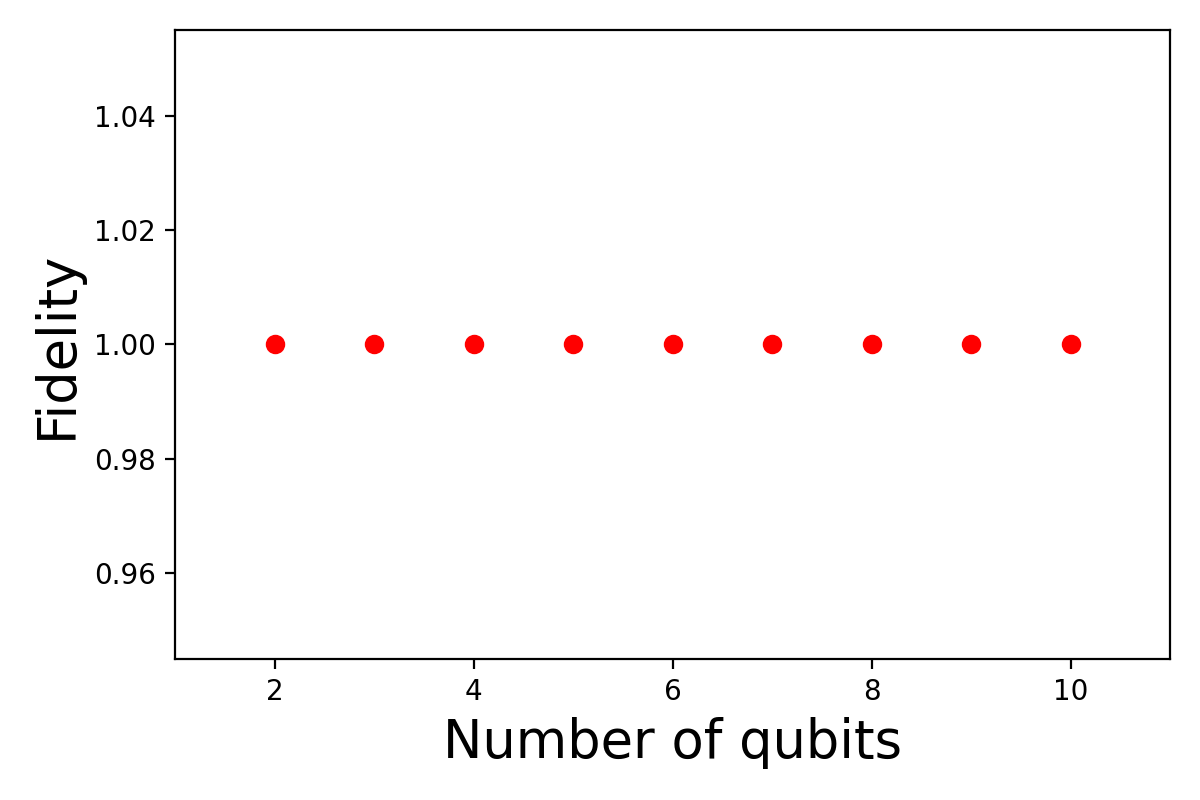}
\caption{Plots of the trace distance and fidelity between the reconstructed density matrix and the true density matrix.}
\label{fig_a3}
\end{figure}

We considered different quantum circuits comprising of 2-10 qubits with random quantum gates as shown in Figure \ref{fig_a3} for one sample 6-qubit circuit, and tried to reconstruct the density matrix followed by its comparison with the true density matrix. The mean measurements necessary for reconstruction of the density matrix are obtained from simulating the quantum circuits. We used the \textit{statevector$\_$simulator} backend in IBM's Qiskit that simulates the quantum circuit without the consideration of errors and noise.
\begin{figure}[ht!]
  \centering 
\includegraphics[width=3.2in]{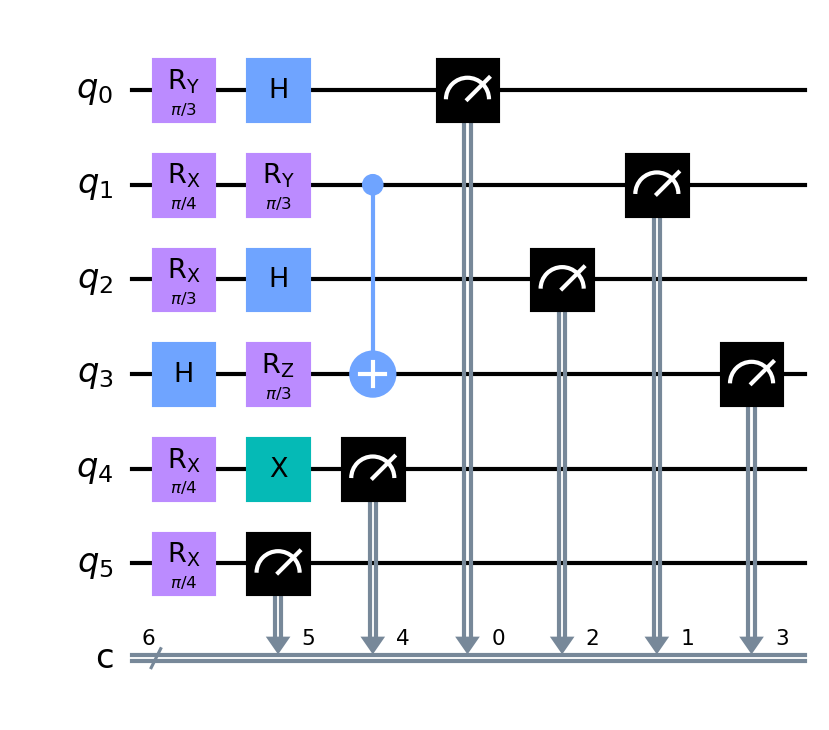} 
\caption{A sample circuit with random quantum gates considered for testing and validating the reconstruction of the density matrix. }
\label{fig_a2}
\end{figure} 
To distinguish between the two states, the trace distance and fidelity between the reconstructed density matrix and true density matrix is calculated for different number of qubits. The results in Figure \ref{fig_a3} show that the states match exactly as the trace distance between the reconstructed state and the true state is zero and the fidelity is one.

\subsection{Convergence to pure state}
\begin{figure}[ht!]
  \centering 
\includegraphics[width=3.2in]{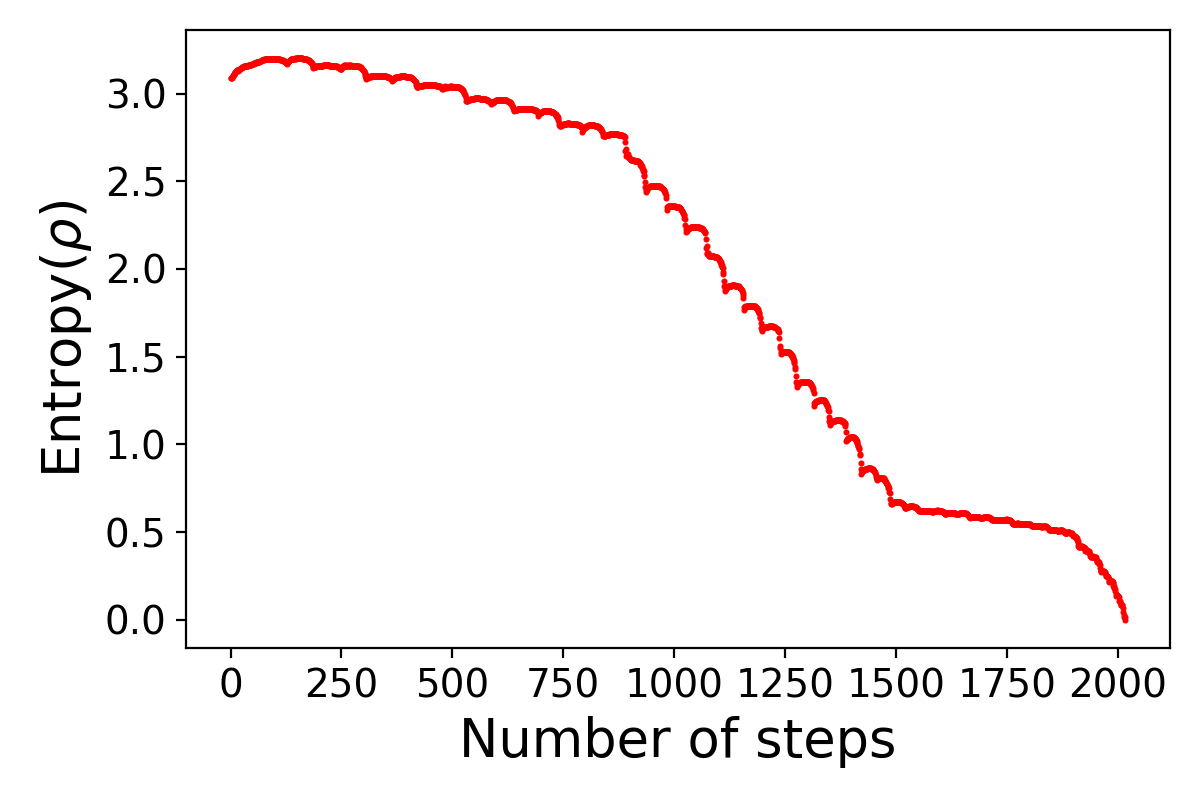} \hspace*{\fill}
\includegraphics[width=3.2in]{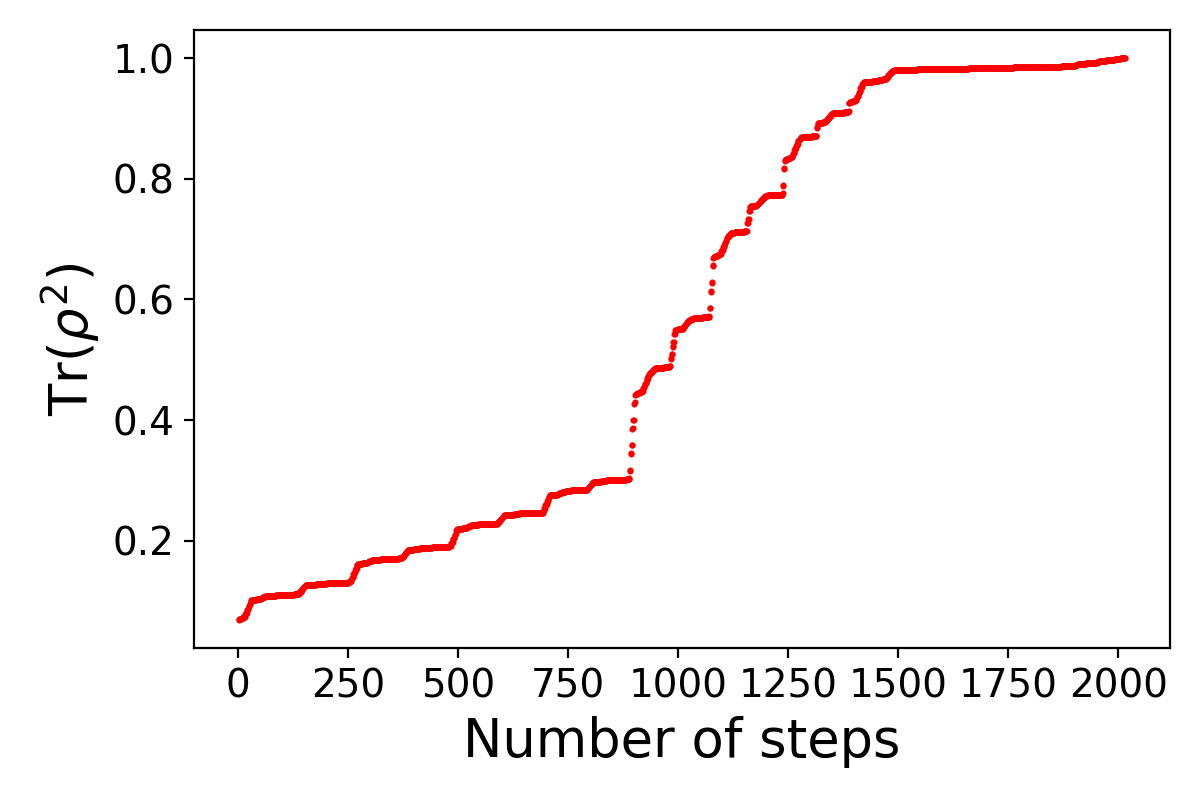} 
\caption{Entropy of the reconstructed density matrix ($\rho$) and trace of $\rho^2$ as a function of number of steps for reconstructing the full density matrix.}
\label{fig_a4}
\end{figure}

Two of the most common parameters to characterize a pure state from its density matrix ($\rho$) are: Entropy($\rho$) = 0 and Tr($\rho^2$) = 1. Our approach is to reconstruct the density matrix of a quantum system using two observables at a time and then repeat the process with different known mean measurements until all the observables have been determined. At each step we have more information about the system which means that we get closer to reconstructing the pure state and therefore, the entropy of the reconstructed density matrix at each step should approach towards zero and the trace of $\rho^2$ should approach towards one. The number of repetitions required for complete reconstruction of \textit{n}-qubit quantum system is \textit{N(N-1)/2} where \textit{N}=$2^n$. Figure \ref{fig_a4} shows the plots of entropy($\rho$) and trace($\rho^2$) at each step for a 6-qubit system for the sample circuit shown in Figure \ref{fig_a2}. The plots validate the convergence of the reconstructed density matrix to a pure state at the end of the complete procedure.   

\section{Error analysis}
\begin{figure}[ht!]
  \centering 
\includegraphics[width=3.2in]{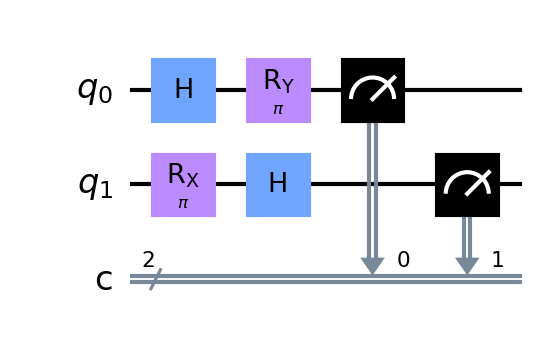} \hspace*{\fill}
\includegraphics[width=3.2in]{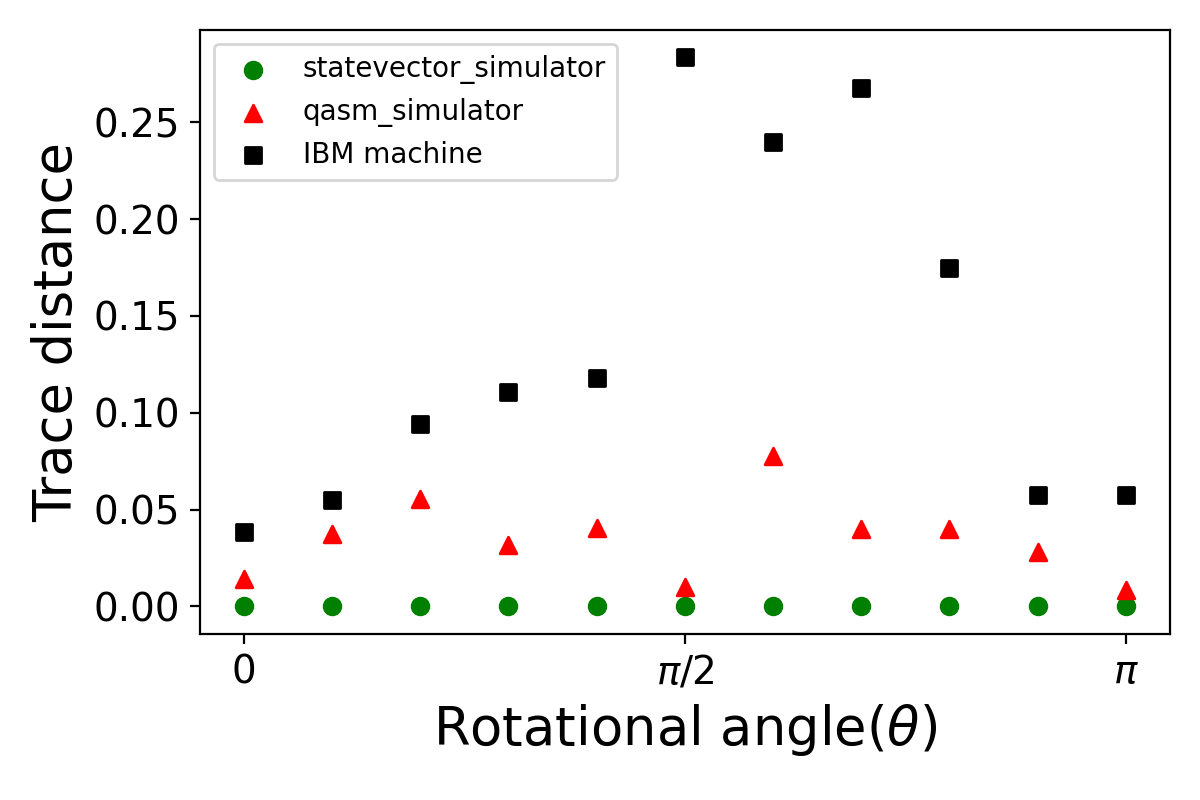} 
\caption{Trace distance between the true and the reconstructed state for the shown quantum circuit as a function of different rotational angles in R$_x$ and R$_y$ gates for the three backends in Qiskit: \textit{statevector$\_$simulator}, \textit{qasm$\_$simulator} and \textit{IBM machine}.}
\label{fig_a5}
\end{figure}

As we saw in the previous section, we considered trace distance, between the reconstructed state and the true state, as a measure to test the performance of our approach. The first step towards the reconstruction of the density matrix is to obtain the mean measurement values of the minimum number of observables required to reproduce the quantum state. These expectation values can be obtained by running the quantum circuit on a quantum computing chip or using an efficient simulator to simulate the expected results. An online platform for cloud-based quantum computing is provided by IBM Quantum Experience on which we conducted numerical experiments to validate our work. \\
The first method that we employed to obtain the mean measurements was the \textit{statevector$\_$simulator} backend in Qiskit in which we can get the final wave function of the simulated state and therefore, the probabilities and coherences can be calculated directly. Since no noise is accounted for in this procedure so the reconstructed state matches exactly with the true state and hence, the trace distance is zero. \\
Next we considered the noisy quantum circuit simulator backend in Qiskit which is the \textit{qasm$\_$simulator} that encompasses statistical errors inversely proportional to the square root of the number of shots given for the circuit to be simulated. As can be seen in Figure \ref{fig_a5} these statistical fluctuations are reflected in the trace distance which is slightly deviated from zero. Also, the operator for calculating coherence unlike probability is not directly available for \textit{qasm$\_$simulator} as well as for the IBM machine and so we decompose the coherence operator into tensor products of Pauli matrices as discussed in [\cite{rishabh},\cite{kandala2017hardware},\cite{bian2019quantum}] and thereby, obtain the coherences. \\
Lastly, we implemented our approach to reconstruct the density matrix for the quantum system corresponding to the state obtained upon running the circuit shown in Figure \ref{fig_a5} on the 5-qubit IBM quantum computing chip: IBM Q 5 Yorktown [\cite{ibm}]. In order to mitigate the measurement errors we also employed the Qiskit’s \textit{ignis.mitigation.measurement} module which does so by constructing a calibration matrix. Considering the presence of noise in the quantum chips the trace distance plot corroborates our approach of carrying out quantum state tomography for a general pure state with any number of qubits.

\section{Concluding remarks}
In this study we have shown the reconstruction of the density matrix of a pure state using expectation values of \textit{N} observables consisting of a probability and \textit{N-1}  coherences. Our approach is based on the formalism of maximal entropy where the constraints for state determination are mean values of populations and coherences. This method provides us with an inference of the quantum state which is then compared with the original state to demonstrate the accuracy of our approach. Our approach focuses on the maximal entropy formalism of the scaled pairwise combinations of the known mean measurements to construct the whole density matrix. It is worth noting that the pairwise maximum entropy approach is invariant under the proposed scaling technique and invites a future fundamental investigation to figure out the origin of this scaling. The simple nature of our proposed formalism can find its application in a variety of different areas where quantum state tomography is essential. We intend to further study this approach to characterize and mitigate quantum gate errors that occur when running circuits on IBM's quantum computing chips. Also, the current approach to reconstruct the density matrix is applied and tested for pure states but it can be a good starting point for the reconstruction of mixed states as well.   

\begin{acknowledgement}
We acknowledge the partial financial support by the U.S. Department of Energy (Office of Basic Energy Sciences) under Award No. DE-SC0019215. We also acknowledge use of the IBM Q for this work. The views expressed here are those of the authors and do not reflect the official policy or position of IBM or the IBM Q team.

\end{acknowledgement}


\bibliography{achemso-demo}

\clearpage
\end{document}





\begin{suppinfo}

\section{Scaling approach for accurate prediction of coherence}
Our approach of reconstructing the density matrix of a quantum system is based on using the pairwise combination of the available known mean measurements to predict an unknown mean measurement. Based on the maximal entropy formalism combined with the method of Lagrange multipliers, given a probability and a coherence, the density operator is given as:
\begin{eqnarray}
\hat{\rho} = \frac{1}{Z(\lambda_{11},\lambda_{12})}\exp\{-\lambda_{11}\ket{1}\bra{1}-\lambda_{12}\ket{1}\bra{2} -\lambda_{12}^{*}\ket{2}\bra{1}\} \label{rho_sc} 
\end{eqnarray}
Since the available operators do not commute so we first diagonalize the exponent term in Eq. (\ref{rho_sc}):
\[
A = 
\begin{bmatrix}
 -\lambda_{11} & -\lambda_{12} & 0 & 0  \\
  -\lambda_{12}^{*} & 0 & 0 & 0 \\
  0 & 0 & 0 & 0 \\
  0 & 0 & 0 & 0 
 \end{bmatrix} =\sum_{i=1}^{4} \epsilon_{i}\ket{\phi_{i}}\bra{\phi_{i}}
\]
with $\{\epsilon_i,\ket{\phi_i}\}$ being eigenvalues and corresponding eigenvectors of the matrix \textbf{A}:
\begin{eqnarray}
\epsilon_1 = 0 &;& \bra{\phi_1} = (0\hspace{0.2cm}0\hspace{0.2cm}0\hspace{0.2cm} 1) \nonumber \\
\epsilon_2 = 0 &;& \bra{\phi_2} = (0\hspace{0.2cm}0\hspace{0.2cm}1\hspace{0.2cm} 0) \nonumber \\
\epsilon_3 = -\frac{1}{2}(\lambda_{11}+\sqrt{4\lambda_{12}\lambda_{12}^{*}+\lambda_{11}^2} ) &;& \bra{\phi_3} = (k_3\hspace{0.2cm}1\hspace{0.2cm}0\hspace{0.2cm} 0) \nonumber \\
\epsilon_4 = -\frac{1}{2}(\lambda_{11}-\sqrt{4\lambda_{12}\lambda_{12}^{*}+\lambda_{11}^2} ) &;& \bra{\phi_4} = (k_4\hspace{0.2cm}1\hspace{0.2cm}0\hspace{0.2cm} 0) \nonumber
\end{eqnarray}
where k$_3$ = -$\frac{\epsilon_3}{\lambda_{12}^{*}}$, k$_4$ = -$\frac{\epsilon_4}{\lambda_{12}^{*}}$ \newline
Now, the density operator based on just two known mean measurements is given as:
\begin{eqnarray}
\hat{\rho} &=& \frac{1}{Z}\big( \exp{\epsilon_1}\ket{\phi_1}\bra{\phi_1}+\exp{\epsilon_2}\ket{\phi_2}\bra{\phi_2}+\exp{\epsilon_3}\ket{\phi_3}\bra{\phi_3}+\exp{\epsilon_4}\ket{\phi_4}\bra{\phi_4} \big) \nonumber \\
Z &=& tr\big( \exp{\textbf{A}} \big) = \sum_{i=1}^4 \exp{\epsilon_i} \nonumber
\end{eqnarray}
Giving the final form of the density operator:
\begin{eqnarray}
\hat{\rho}&=&\frac{1}{Z}\sum_{i}\exp{\epsilon_{i}}|\phi_{i}><\phi_{i}| \nonumber \\
&=&\frac{1}{Z}(|4><4|+|3><3|+(a+b)|1><1|+(\frac{a}{k_{3}^*}+\frac{b}{k_{4}^*})|1><2|  \nonumber\\
&+&(\frac{a}{k_{3}}+\frac{b}{k_{4}})|2><1|+(\frac{a}{\abs{k_{3}}^{2}}+\frac{b}{\abs{k_{4}}^{2}})|2><2|) \label{density}
\end{eqnarray}
where Z=$\sum_{i}\exp{\epsilon_{i}}$, k$_3$ = -$\frac{\epsilon_3}{\lambda_{12}^{*}}$, k$_4$ = -$\frac{\epsilon_4}{\lambda_{12}^{*}}$, a=$\frac{\abs{k_{3}}^{2}}{\sqrt{(k^2_3+1)({k_3^*}^2+1)}}\exp{\epsilon_{3}}$, and \newline b=$\frac{\abs{k_{4}}^{2}}{\sqrt{(k^2_4+1)({k_4^*}^2+1)}}\exp{\epsilon_{4}}$. \\
The unknown mean measurement value is determined using the coefficient of the corresponding operator in Eq. (\ref{density}). 
If we consider the maximally entangled state: $\ket{\psi}$ = $\frac{1}{\sqrt{N}}(\ket{00}+\alpha\ket{11})$ and try to predict the coherence terms based on the above approach we get very accurate results for all values of $\alpha$ as shown in Figure \ref{fig_ap1}.
\begin{figure}[ht!]
  \centering 
  \includegraphics[width=3.2in]{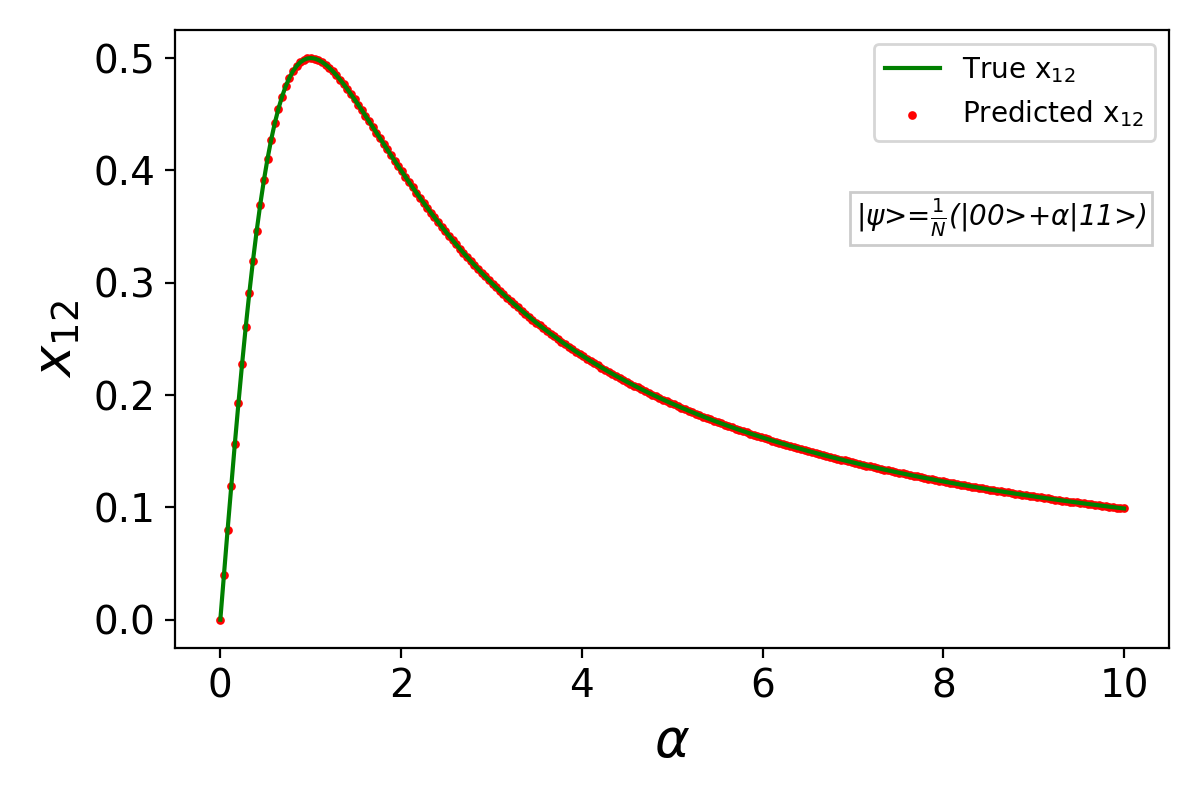}
\caption{Prediction of coherence x$_{12}$ for a general 2-qubit entangled state: $\ket{\psi}$ = $\frac{1}{\sqrt{N}}(\ket{00}+\alpha\ket{11})$ using the maximal entropy formalism.}
\label{fig_ap1}
\end{figure}
However, we know from Figure \ref{fig_a1}a that if more than 2 non-zero coefficients of basis states are present, the accuracy of prediction of coherence falls for smaller coefficients. This is because in such cases the determined value of Lagrange multiplier $\lambda_{12}$ is very small. This leads to numerical instability in the calculation of the projection operator $\ket{\phi_i}\bra{\phi_i}$ where $\lambda_{12}$ comes in the denominator. Seeking motivation from the accuracy of the result of density matrix reconstruction for the Bell state and also accounting for the fact that in our approach we consider two known mean measurements at a time we propose that we can scale the probabilities and the unknown coherence such that the probabilities correspond to the probabilities of a two level system i.e. the sum of the probabilities is one. 
\begin{figure}[ht!]
  \centering 
  \subcaptionbox{Prediction of x$_{12}$ using maximal entropy formalism}
{\includegraphics[width=3.2in]{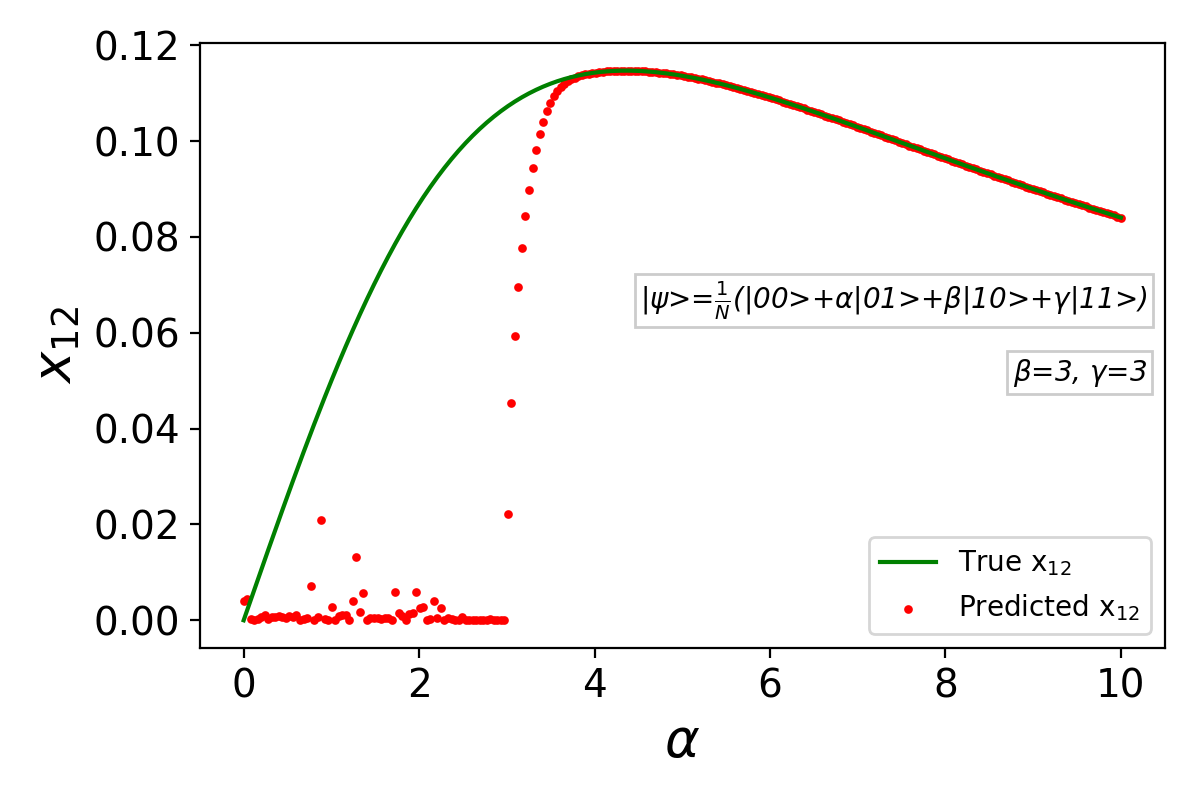}} \hspace*{\fill}
\subcaptionbox{Prediction of x$_{12}$ using maximal entropy formalism combined with scaling technique}
{\includegraphics[width=3.2in]{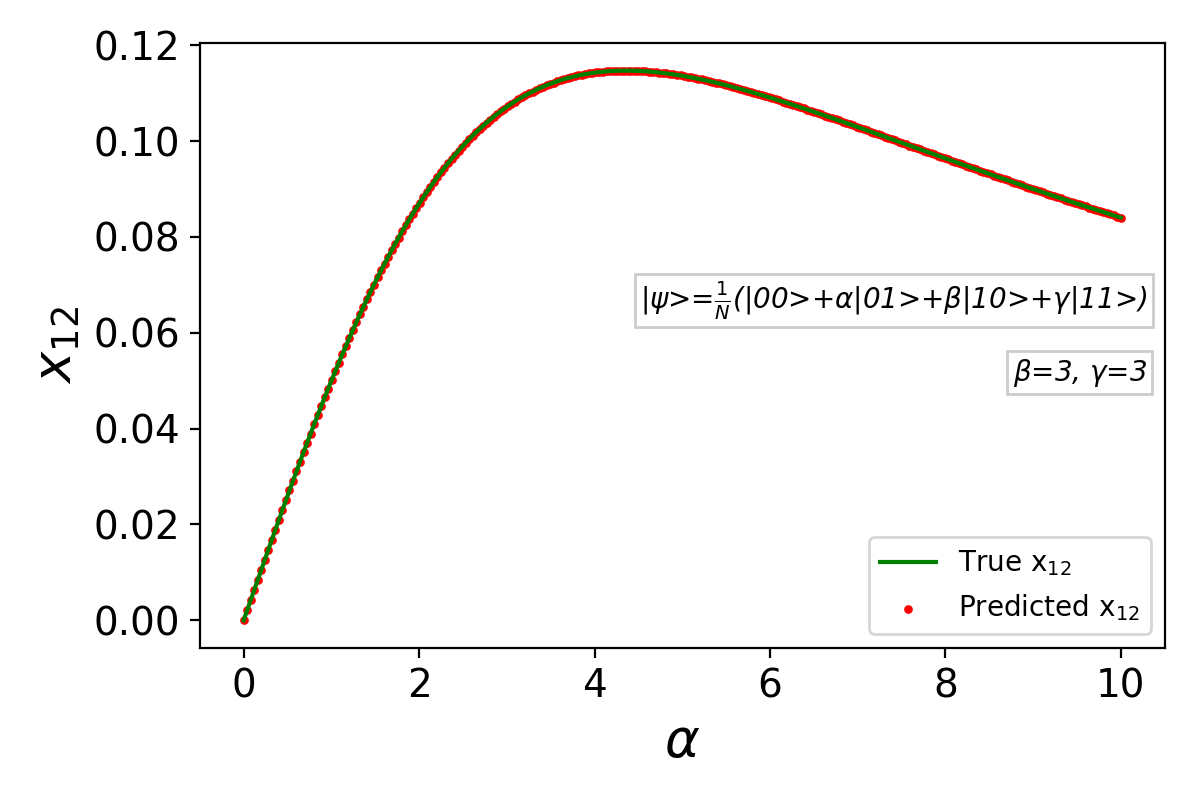}}
\caption{Plot of the predicted and true mean measurement x$_{12}$ versus the coefficient $\alpha$ of the state $\ket{01}$ using the maximal entropy formalism from: (a) two known probabilities x$_{11}$ and x$_{22}$ (b) two known probabilities x$_{11}$ and x$_{22}$ and also employing the scaling approach.  }
\label{fig_a1}
\end{figure}
For example, if we want to determine x$_{12}$ from x$_{11}$ and x$_{22}$ but the usual approach does not predict an accurate coherence then we carry out the scaling as:
    \[
\begin{bmatrix}
  x_{11} & x_{12}   \\
  x_{12}^{*} & x_{22}  
 \end{bmatrix} 
 \longrightarrow
 s
 \begin{bmatrix}
  x_{11} & x_{12}   \\
  x_{12}^{*} & x_{22}  
 \end{bmatrix} =
  \begin{bmatrix}
  sx_{11} & sx_{12}   \\
  sx_{12}^{*} & sx_{22}  
 \end{bmatrix} =
  \begin{bmatrix}
  x_{11}^{\prime} & x_{12}^{\prime}   \\
  x_{12}^{*\prime} & x_{22}^{\prime}  
 \end{bmatrix}
 \hspace{0.7cm} (such \hspace{0.2cm} that \hspace{0.2cm} x_{11}^\prime + x_{22}^{\prime} = 1)
\]
We apply the maximal entropy formalism to calculate the coherence x$_{12}^{\prime}$ which is accurate and then back scale the probabilities and coherence:
\[
\begin{bmatrix}
  x_{11} & x_{12}   \\
  x_{12}^{*} & x_{22}  
 \end{bmatrix} =
 \frac{1}{s}
  \begin{bmatrix}
  x_{11}^{\prime} & x_{12}^{\prime}   \\
  x_{12}^{*\prime} & x_{22}^{\prime}  
 \end{bmatrix}
\]
The maximal entropy approach on the pairwise combination of the known mean measurements is invariant under the application of the proposed scaling technique. The scaling method is implemented to avoid the numerical singularity that is resulted when the Lagrange multipliers are close to zero. An initial study of the eigenvalues of the density matrix before and after scaling shows that the eigenvalues of the initial density matrix without scaling are all zero except for one eigenvalue which is one. After using the scaling technique, two of the eigenvalues are zero and the remaining are non-zero for the density matrix of a 2-qubit quantum system. However, since the accuracy of prediction is excellent so a further investigation is going on to figure out the origin of this scaling technique at a fundamental level. \\
To verify this approach, we considered the maximally entangled state for 3-qubits: $\ket{\psi}$ = $\frac{1}{\sqrt{8}}$($\ket{000}+\ket{001}+\ket{010}+\ldots+\ket{111}$). We reconstructed the density matrix, first by using the maximal entropy approach without the use of the scaling technique as mentioned above and obtained the following density matrix:
\[
\rho_{rec} =  
\begin{bmatrix}
  0.125 & 0 & 0 & 0 & 0 & 0 & 0 & 0  \\
  0 & 0.125 & 0 & 0 & 0 & 0 & 0 & 0 \\
  0 & 0 & 0.125 & 0 & 0 & 0 & 0 & 0 \\
  0 & 0 & 0 & 0.125 & 0 & 0 & 0 & 0 \\
  0 & 0 & 0 & 0 & 0.125 & 0 & 0 & 0 \\
  0 & 0 & 0 & 0 & 0 & 0.125 & 0 & 0 \\
  0 & 0 & 0 & 0 & 0 & 0 & 0.125 & 0 \\
  0 & 0 & 0 & 0 & 0 & 0 & 0 & 0.125 
 \end{bmatrix} 
\]
All the coherences turned out to be zero as the determined Lagrange multipliers in each case were zero. However, if we use the above scaling technique and combine it with the maximal entropy approach, we do get an accurate reconstructed density matrix:
\[
\rho_{rec}^{scaled} =  
\begin{bmatrix}
  0.125 & 0.125 & 0.125 & 0.125 & 0.125 & 0.125 & 0.125 & 0.125  \\
  0.125 & 0.125 & 0.125 & 0.125 & 0.125 & 0.125 & 0.125 & 0.125  \\
  0.125 & 0.125 & 0.125 & 0.125 & 0.125 & 0.125 & 0.125 & 0.125  \\
  0.125 & 0.125 & 0.125 & 0.125 & 0.125 & 0.125 & 0.125 & 0.125  \\
  0.125 & 0.125 & 0.125 & 0.125 & 0.125 & 0.125 & 0.125 & 0.125  \\
  0.125 & 0.125 & 0.125 & 0.125 & 0.125 & 0.125 & 0.125 & 0.125  \\
  0.125 & 0.125 & 0.125 & 0.125 & 0.125 & 0.125 & 0.125 & 0.125  \\
  0.125 & 0.125 & 0.125 & 0.125 & 0.125 & 0.125 & 0.125 & 0.125  
 \end{bmatrix} 
\]
Therefore, we can use this method to reconstruct the full density matrix for any general pure state with real amplitudes and for any number of qubits using only \textit{N} probabilities. In case of complex coherences, this approach can be used to determine the amplitude of the coherence that can then be combined with the phase estimation algorithm as discussed in the main body of the paper. Thus, using the mean measurement values of \textit{N} observables we can reconstruct the full density matrix of a quantum system in pure state using this approach.

\end{suppinfo}
